\documentclass[doublecol,a4paper,showpacs]{epl2}
\usepackage{tensor}
\usepackage{graphicx}
\usepackage{amsmath}
\usepackage{amssymb}
\usepackage{enumerate}
\usepackage{subfigure}
\usepackage{tabularx}
\usepackage[colorlinks=true, pdfstartview=FitV, linkcolor=blue, citecolor=red, urlcolor=black, breaklinks=true]{hyperref}
\newcommand{\be}{\begin{equation}}
\newcommand{\ee}{\end{equation}}
\newcommand{\ben}{\begin{eqnarray}}
\newcommand{\een}{\end{eqnarray}}
\newcommand{\bes}{\begin{subequations}}
\newcommand{\ees}{\end{subequations}}
\def\bal#1\eal{\begin{align}#1\end{align}}
\newcommand{\vphi}{\varphi}

\newcommand{\LL}{{\mathcal L}}

\begin{document}
\title{Quasi-compact vortices}
\author{D. Bazeia\inst{1} \and M.A. Marques\inst{1} \and R. Menezes\inst{2,1}}
\shortauthor{D. Bazeia \etal}

\institute{                    
  \inst{1} Departamento de F\'\i sica, Universidade Federal da Para\'\i ba, 58051-970 Jo\~ao Pessoa, PB, Brazil\\
  \inst{2} Departamento de Ci\^encias Exatas, Universidade Federal da Para\'{\i}ba, 58297-000 Rio Tinto, PB, Brazil}

\abstract{
We deal with planar vortex structures in Maxwell-Higgs models in the presence of a generalized magnetic permeability. The model under investigation engenders a real parameter that controls the behavior of the tail of the solutions and of the quantities associated to them. As the parameter gets larger, the solutions attain their boundary values faster, unveiling the existence of a peculiar feature, the presence of double exponential tails. However, the solutions are not compact so we call them \emph{quasi}-compact vortices.}
\maketitle

The study of defect structures such as kinks and vortices has been of direct interest to high energy physics \cite{vilenkin,manton,vachaspati,weinberg}. Among the diverse applications, they may be useful not only in high energy physics, but also in condensed matter \cite{fradkin,hubert}, where they describe specific features associated to magnetic materials, for instance. The asymptotic behavior of the solutions that describe topological objects is important because it is associated to the force between them and their scattering. The tail of the solution also plays a role in formation of structures and other features \cite{collision}. In the case of kinks, the standard configurations usually show an asymptotic behavior given by an exponential function of the spatial coordinate. Recently, several investigations dealing with kinks that present a non exponential asymptotic behavior has appeared in the literature \cite{long1,long2,long3,long4,long5,comp1,comp2,comp3,comp4}. In Refs.~\cite{long1,long2,long3,long4,long5}, the authors have studied kinks that exhibit an asymptotic profile of the polynomial type, in which the kinks presents an interesting collective demeanor, with a long range interaction. On the other hand, in Refs.~\cite{comp1,comp2,comp3,comp4}, the authors investigated kinks that engender a compact profile, attaining the boundary conditions in a finite interval of the line, so the solution does not present a tail; this makes the interaction between the structures separated by a distance larger than the sum of their widths vanish.

As one knows, in order to study vortices, one has to deal with two fields that are coupled through an $U(1)$ symmetry \cite{NO}: a complex scalar and a gauge field. In this scenario, the standard solutions and the quantities associated to them, such as the magnetic field and the energy density, exhibit an asymptotic profile given by the product of a polynomial with an exponential function of the radial coordinate. In Refs.~\cite{compvortex,compcs}, the authors introduced models that support vortices with a compact profile. The models engender a parameter that controls the asymptotic behavior of the solutions. As the parameter increases, the tail of the solutions shrinks to accommodate inside a compact space. Furthermore, depending on the specific model, the magnetic field compactifies into a disk \cite{compvortex} or ring \cite{compcs}.

In this letter, in the Maxwell-Higgs scenario with a generalized magnetic permeability, we investigate the presence of solutions  that exhibit double exponential tails, which we call \emph{quasi}-compact vortices. The investigation is inspired by
    Ref. \cite{qballs}, in which we studied quasi-compact Q-balls. In the present work, however, we consider $(2,1)$ spacetime dimensions and the Lagrange density given by
\be\label{lmodel}
	\LL = - \frac{1}{4\mu(|\vphi|)}F_{\alpha\beta}F^{\alpha\beta} +|D_\alpha\vphi|^2 - V(|\vphi|).
\ee
Here, we work with dimensionless fields and coordinates. The above model describes two fields: $\vphi$, which is a complex scalar field, and $A_\alpha$, a gauge field; they are coupled through the $U(1)$ local symmetry. The electromagnetic tensor is $F_{\alpha\beta}=\partial_{\alpha}A_\beta-\partial_{\beta}A_\alpha$ and the covariant derivative has the form $D_\alpha=\partial_{\alpha}+iA_{\alpha}$. The function $V(|\vphi|)$ denotes the potential and $\mu(|\vphi|)$ stands for a generalized magnetic permeability. We suppose $\mu(|\vphi|)$ is nonnegative and take the metric tensor $\eta_{\alpha\beta}=(1,-1,-1)$. It is straightforward to see that the Nielsen-Olesen model \cite{NO,vega} is recovered for $\mu(|\vphi|)=1$. 

The Lagrange density \eqref{lmodel} has the following associated equations of motion
\bes\label{geom}
\begin{align}
& D_\alpha D^\alpha \vphi + \frac{\vphi}{2|\vphi|}\!\left(-\frac{\mu_{|\vphi|}}{4\mu^2} F_{\alpha\beta}F^{\alpha\beta} + V_{|\vphi|} \right)\! =0,\\ \label{meqsc}
& \partial_\alpha \left(\frac1\mu F^{\alpha\beta}\right) = J^\beta.
\end{align}
\ees
We are using $\mu_{|\vphi|}=d\mu/d|\vphi|$ and $V_{|\vphi|} = \partial V/\partial{|\vphi|}$, and in the latter expression, we have $J_{\alpha} = i(\overline{\vphi} D_{\alpha} \vphi-\vphi\overline{D_{\alpha}\vphi})$ .

To investigate the presence of vortices, we consider static configurations. We also take $A_0=0$, which leads to electrically neutral structures as one can show by setting $\nu=0$ in Eq.~\eqref{meqsc}. For the other functions, the following ansatz is considered:
\be\label{ansatz}
\vphi = g(r)e^{in\theta} \quad\text{and}\quad \vec{A} = {\frac{\hat{\theta}}{r}\left(n-a(r)\right)},
\ee
where $n$ is an integer number that represents the vorticity. In this case, $a(r)$ and $g(r)$ obey the boundary conditions
\bes\label{bc}
\bal
g(0) &=0, & a(0)&=n,\\
g(\infty)&=1, & a(\infty)&=0,
\eal
\ees
One can use the definition of magnetic field, $B = -F^{12}$, to show that, combined with the ansatz \eqref{ansatz}, it takes the form
\be\label{b}
B = -\frac{a^\prime}{r},
\ee
where the prime represents derivation with respect to the radial coordinate. The specific form of the magnetic field in the above equation allows us to show that its flux is quantized
\be\label{mflux}
\Phi =2\pi\int_0^\infty rdr B = 2\pi n.
\ee
The ansatz \eqref{ansatz} transforms the equations of motion for our fields into
\bes\label{secansatz}
\begin{align}
\frac{1}{r} \left(r g^\prime\right)^\prime &= \frac{a^2g}{r^2} - \frac{\mu_g{a^\prime}^2}{4\mu^2r^2} + \frac12 V_g, \\
r\left(\frac{a^\prime}{\mu r}\right)^\prime &= 2ag^2.
\end{align}
\ees
These equations of motion are nonlinear second order differential equations that couple $a(r)$ and $g(r)$. So, they are very hard to be solved. To simplify the problem, let us search for first order differential equations that are compatible with the above equations of motion. As one knows, the invariance of the Lagrange density in Eq.~\eqref{lmodel} over spacetime translations lead to an energy momentum tensor, which we denote by $T_{\alpha\beta}$. In particular, the energy density is given by $\rho\equiv T_{00}$. One can show that, with the ansatz \eqref{ansatz}, the energy density is given by
\be\label{rhoans}
\rho =\frac{{a^\prime}^2}{2\mu(g)\, r^2} + {g^\prime}^2 +\frac{a^2g^2}{r^2} + V(g).
\ee
Notice that the condition $\mu(|\vphi|)\geq0$ in the interval of existence of the solution contribute to avoid the presence of negative energies. To find the aforementioned first order equations, one can follow Refs.~\cite{leenam,bazeia1992,dmitry}, which develops the Bogomol'nyi procedure \cite{bogo,atmaja}, writing the above expression as a sum of squared terms and a total derivative. The formalism requires the potential to be
\be\label{potgen}
V(|\vphi|) =  \frac{\mu(|\vphi|)}{2}\left(1-|\vphi|^2\right)^2.
\ee
The first order equations arise in the case of minimum energy, given by $E=2\pi|n|$, in which the functions $a(r)$ and $g(r)$ must obey
\bes\label{fovortex}
\bal\label{fog}
g^\prime &= \pm\frac{ag}{r}, \\ \label{foa}
-\frac{a^\prime}{r} &=\pm \mu(g)\left(1-g^2\right).
\eal
\ees
The pair of equations for the lower signs (negative vorticity) are related to the one for the upper signs (positive vorticity) by the change $a\to-a$. For simplicity, we only work with the upper signs and $n=1$. The presence of these first order equations also ensures the stability of the solutions under rescaling; see Ref.~\cite{godvortex}. We see that the generalized magnetic permeability appears in the above first order equations and in the potential \eqref{potgen}. One must be careful when chosing it, since it must lead to potentials with proper behavior and finite energy configurations whose solutions are compatible with the boundary conditions in Eq.~\eqref{bc}. The above first order equations also allow us to write
\be\label{rhov}
\rho = \frac{{a^\prime}^2}{\mu(g)\, r^2} + 2{g^\prime}^2.
\ee

As we have commented before, the standard first order vortex equations, investigated by Bogomol'nyi in Ref.~\cite{bogo}, is recovered for $\mu(|\vphi|)=1$. In this case, the potential in Eq.~\eqref{potgen} has the form
\be
V_s(|\vphi|) =  \frac{1}{2}\left(1-|\vphi|^2\right)^2
\ee
and the first order equations \eqref{fovortex} become
\bes\label{fono}
\bal\label{fogno}
g_s^\prime &= \frac{a_sg_s}{r}, \\ \label{foano}
-\frac{a_s^\prime}{r} &= 1-g_s^2,
\eal
\ees
where the $s$ index stands for the standard solutions. The analytical solutions of the above equations remain unknown. Thus, one must use a numerical approach to calculate them. The result can be found in Ref.~\cite{vega}, so we omit further details here. The asymptotic behavior can be estimated by taking $a(r)=0+a_{asy}(r)$ and $g(r) = 1-g_{asy}(r)$ in the above first order equations up to linear terms in $a_{asy}$ and $g_{asy}$. Here, we obtain
\be\label{asymptno}
a_{asy} = \lambda\,\sqrt{r}\,e^{-\sqrt{2}\,r}\quad\text{and}\quad g_{asy} = \lambda\,\frac{e^{-\sqrt{2}\,r}}{\sqrt{2\,r}},
\ee
where $\lambda$ is an integration constant that can be used to fit the solutions found through numerical simulations. Both solutions vanish far from the origin. Even though they present an exponential factor, we see that $g_{asy}(r)$ vanishes faster than $a_{asy}(r)$ due to the presence of a power of the radial coordinate in the denominator. We remark here that the profile of these solutions are different from the ones for kinks in $(1,1)$ spatial dimensions, in which the asymptotic behavior is controlled by an exponential, without polynomial factors of the spatial coordinate.

We can also use $\mu(|\vphi|)=|\vphi|^2$. In this case, the potential in Eq.~\eqref{potgen} becomes a $|\vphi|^6$ potential, and the first order equations \eqref{fovortex} reproduce the very same first order equations of the Chern-Simons model \cite{jackiw,coreanos}.

We further notice that the presence of the generalized magnetic permeability $\mu(|\vphi|)$ changes the asymptotic behavior of the solutions. By taking a similar approach as in the standard case, one can show that, in the general case, Eq.~\eqref{asymptno} changes to
\be\label{asympt}
a_{asy} = \lambda\,\sqrt{r}\,e^{-\sqrt{2\mu(1)}\,r}\quad\text{and}\quad g_{asy} = \lambda\,\frac{e^{-\sqrt{2\mu(1)}\,r}}{\sqrt{2\mu(1)\,r}},
\ee
where $\lambda$ is an integration constant. This expression works only for finite $\mu(1)$, such that $\mu(1)>0$. One must be careful with the possibility of including infinite and null magnetic permeabilities, which must be treated separately, in a  case-by-case investigation.

By observing the form of $a_{asy}(r)$ and $g_{asy}(r)$ in Eq.~\eqref{asympt}, we notice that $\mu(1)$ controls the asymptotic behavior of the solution: if $\mu(1)$ gets larger, the tail of the solutions becomes smaller. This property was used in Ref.~\cite{compvortex} to compactify the vortex, in particular with
\be\label{muk}
\mu_k(|\vphi|) = \frac{1-|\vphi|^{2k}}{1-|\vphi|^2},
\ee 
where $k$ is a non negative real parameter. In this case, the first order equations \eqref{fovortex} become
\bes\bal
g^\prime &= \frac{ag}{r} \\
 -\frac{a^\prime}{r} &= 1-g^{2k}.
\eal
\ees
We can see that, as $k$ increases, the tail of each solution shrinks more and more, because $\mu(1)=k$. In the limit $k\to\infty$, the solutions become compact, with the analytical form
\bes\label{solc}
\ben
a_c(r)&=&
\begin{cases}
1-\frac{1}{2}r^2,\,\,\,&r\leq \sqrt{2},\\
0, \,\,\, & r>\sqrt{2}.
\end{cases} \\
g_c(r)&=&
\begin{cases}
\frac{r}{\sqrt{2}}\, e^{(2-r^2)/4},\,\,&r\leq \sqrt{2},\\
1, \,\, & r>\sqrt{2}.
\end{cases}
\een
\ees
Here, the $c$ index denotes the compact profile. Both the energy density and the magnetic field associated to these solutions are compact. An interesting feature, in particular, is that the magnetic field is uniform inside the compact space. Since vortices become strings when studied in three spatial dimensions, this behavior maps the magnetic field of an infinitely long solenoid.

 Next, we use a similar argument to find an unprecedented feature regarding the asymptotic behavior of the vortex: the double exponential tail. We first introduce model whose generalized magnetic permeability is controlled by the function
\be\label{mul}
\mu_l(|\vphi|)=l^2\!\left(1-\left|1-|\vphi|^{2}\right|^{2/l}\right)^2,
\ee
where $l$ is a non-negative real parameter. Here, we have $\mu(1)=l^2$, which is associated to the asymptotic demeanor of the solutions. From Eq.~\eqref{asympt}, we get that, as $l$ increases, both $g(r)$ and $a(r)$ attain their boundary values faster, with their tail shrinking more and more. One may think that this behavior leads to a compact profile as in the case described by the magnetic permeability in Eq.~\eqref{muk}. Here, however, we get a distinct behavior, associated to solutions with double exponential tails, which we call \emph{quasi}-compact vortices, since they engender a tail suppression that is much stronger than the standard vortex (that presents exponential tail) and arise in a model similar to the one that supports compact vortices \cite{compvortex} and also, to another one in which we investigated \emph{quasi}-compact Q-balls \cite{qballs}.

To see how this works explicitly, we notice that the limit $l\to\infty$ is special; indeed, one can show that the magnetic permeability in Eq.~\eqref{mul} takes the form
\be
\mu_\infty(|\vphi|) = \ln^2\!\left(1-|\vphi|^2\right)^2.
\ee
In this case, the asymptotic behavior of the solutions admit the form
\be\label{dexp}
g(r)\approx 1- \frac12e^{-\lambda e^{\sigma r}}\quad\text{and}\quad a(r) \approx \frac{\lambda\sigma}{2}re^{\sigma r}e^{-\lambda e^{\sigma\,r}},
\ee
with $\sigma=2\sqrt{2}$. Here, $\lambda$ is an integration constant that can be used to fit the curves associated to these expressions with the numerical simulation. The above expressions show that the solutions engender a double exponential tail. Therefore, the solutions go faster to their boundary values as $l$ increases, with their tail approaching to above double exponential profile. They arise in the limit $l\to\infty$, with infinite $\mu_\infty(1)$, but are not compact as in the case described by Eq.~\eqref{muk} with $k\to\infty$; so, we call them \emph{quasi}-compact solutions. By remembering that the polynomial $\left(1-e^r/l\right)^l$ tends to $e^{-e^r}$ for $l\to\infty$, we can see there is no way to make a connection between the double exponential and the expressions in Eq.~\eqref{asympt}, which are the lowest order approximations of $e^{-r}$, because we need the contributions in all orders of exponentials to do so.

\begin{figure}[t!]
		\centering
		\includegraphics[width=8.5cm,trim={0.65cm 1cm 0 0},clip]{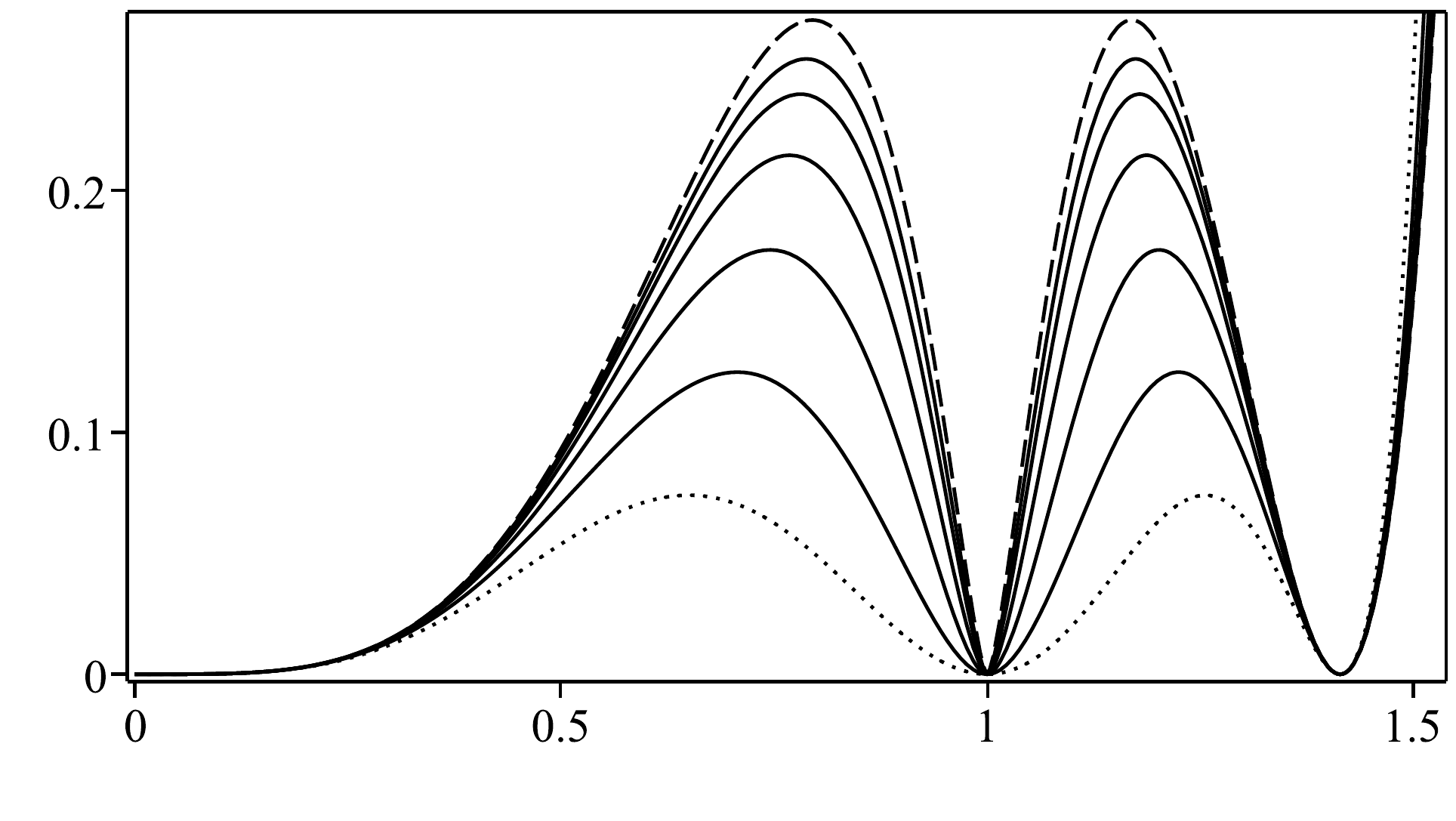}
		\caption{The potential in Eq.~\eqref{potl} for $l=1$ (dotted line), $2,\,4,\,8,\,16,\,32$ and the limit $l\to\infty$ in Eq.~\eqref{potlog} (dashed line).}
		\label{figpot}
		\end{figure}

Considering the form of the magnetic permeability given by Eq.~\eqref{mul}, the potential in Eq.~\eqref{potgen} becomes
\be\label{potl}
V_l(|\vphi|) =  \frac{l^2}{2}\left(1-|\vphi|^2\right)^2\left(1-\left|1-|\vphi|^{2}\right|^{2/l}\right)^2.
\ee
In the limit $l\to\infty$, we get
\be\label{potlog}
V_\infty(|\vphi|) =  \frac{1}{2}\left(1-|\vphi|^2\right)^2\ln^2\!\left|1-|\vphi|^2\right|^2.
\ee
In Fig.~\ref{figpot}, we display the potential for several values of $l$, including the above limit. We are interested in the interval $|\vphi|\in[0,1]$, which is the one where the vortex exists. As one can see in the figure, $|\vphi|=0$ and $|\vphi|=1$ are minima of the potential connected by the scalar field solution and $|\vphi_m|^2 = 1-\left(l/(l+4)\right)^{l/2}$ is a set of maximum points whose limit for $l\to\infty$ is $|\vphi_m|=\sqrt{1-1/e^2}$.

\begin{figure}[t!]
		\centering
		\includegraphics[width=8.5cm,trim={0.65cm 1cm 0 0},clip]{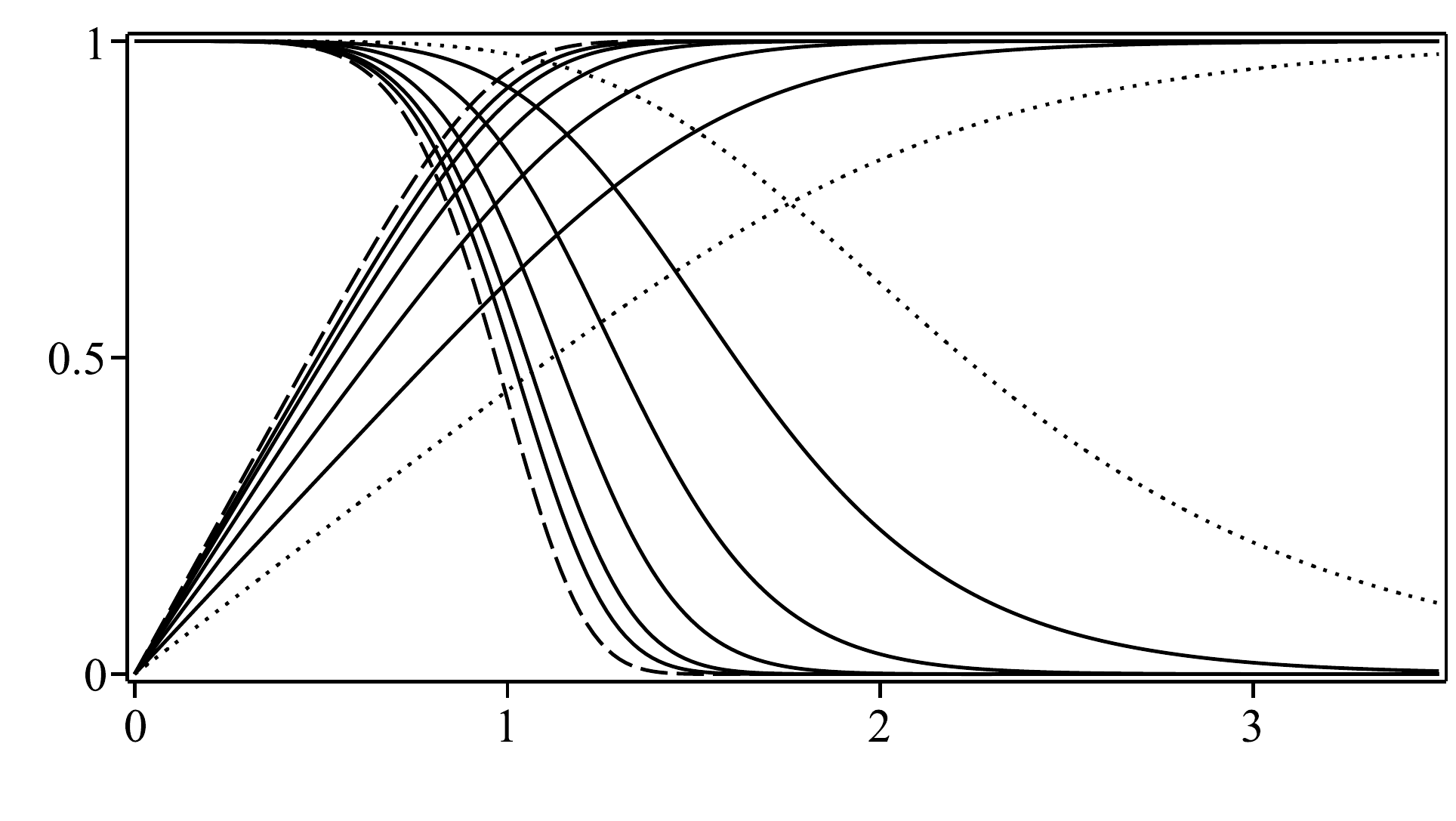}
		\caption{The vortex solutions $a(r)$ (ascending lines) and $g(r)$ (descending lines) for $l=1$ (dotted lines), $2,\,4,\,8,\,16,\,32$ and the limit $l\to\infty$ (dashed lines).}
		\label{figsol}
		\end{figure}

The first order equation \eqref{fog} does not depend on $\mu(g)$. We must solve it combined with Eq.~\eqref{foa}, which becomes
\be\label{fol}
-\frac{a^\prime}{r} = l^2 (1-g^2)\left(1-\left(1-g^{2}\right)^{2/l}\right)^2.
\ee
In the limit $l\to\infty$, we get
\be\label{foinf}
-\frac{a^\prime}{r} = (1-g^2)\ln^2\!\left(1-g^2\right)^2.
\ee
We have not been able to find analytical solutions of the above equations, so we used numerical procedures to calculate the solutions, which are displayed in Fig.~\ref{figsol}. The solution $g(r)$, associated to the scalar field, connects the minimum $|\vphi|=0$ and the set of minima described by $|\vphi|=1$. As we can see, as $l$ increases, the solutions attain their boundary values faster. At first glance, one may think the solutions are compactifying as in Refs.~\cite{compvortex,compcs,anavortex}. Nevertheless, as we have shown in Eq.~\eqref{dexp}, the dashed solutions are not compact: they engender a double exponential tail. In order to highlight this feature, we call the structure a \emph{quasi}-compact vortex.
		\begin{figure}[b!]
		\centering
		\includegraphics[width=8.5cm,trim={0.65cm 1cm 0 0},clip]{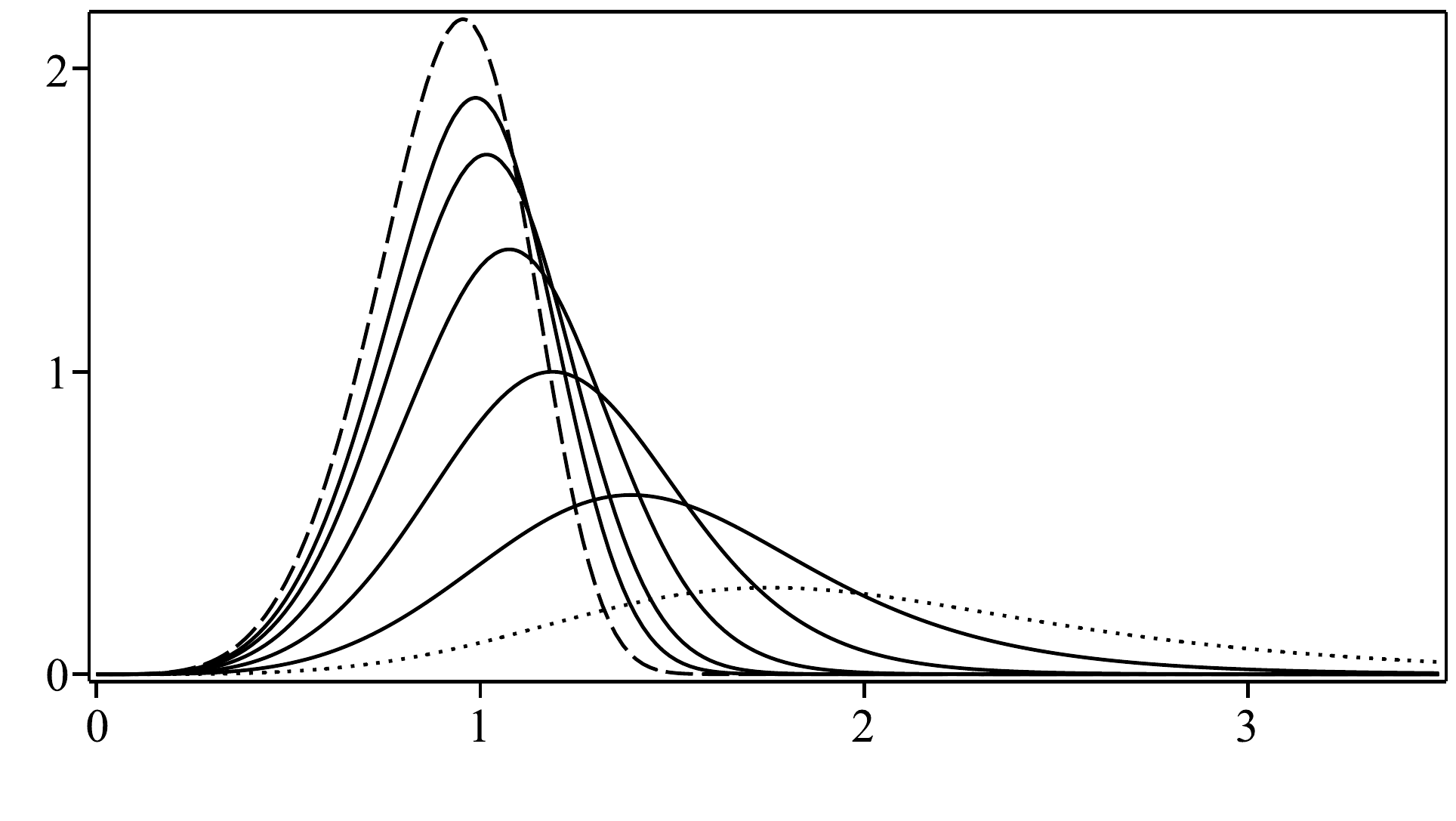}
		\includegraphics[width=4.25cm,trim={0.5cm 0.5cm 0.5cm -0.5cm},clip]{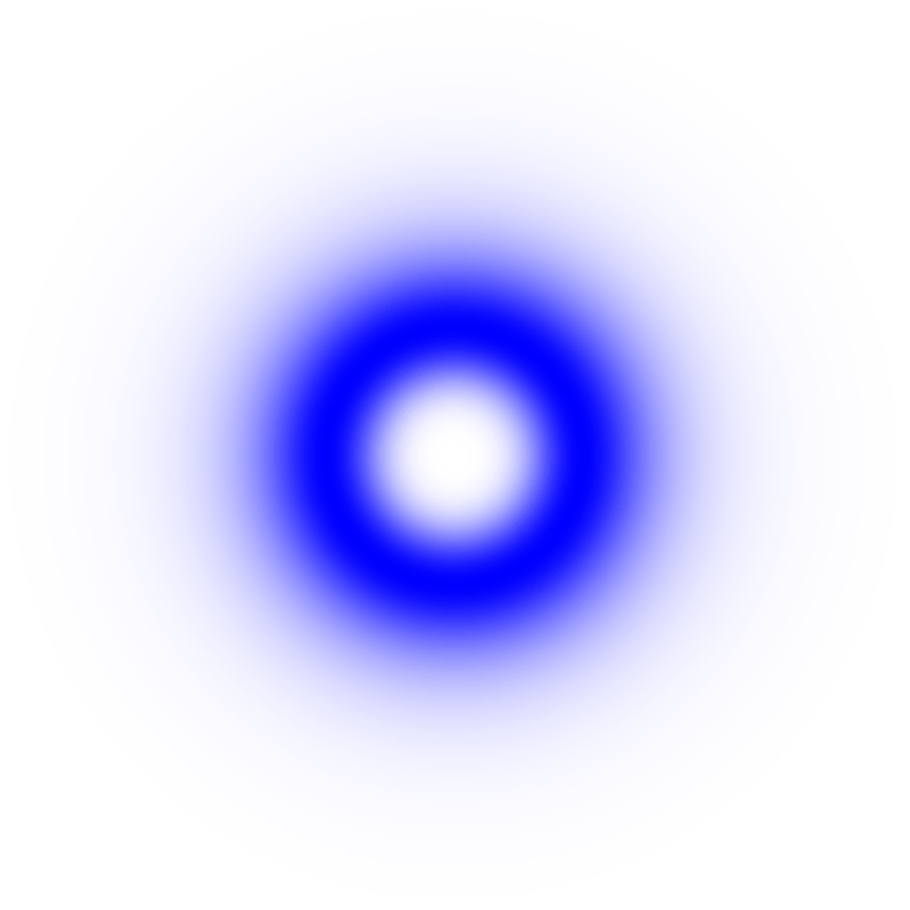}
		\includegraphics[width=4.25cm,trim={0.5cm 0.5cm 0.5cm -0.50cm},clip]{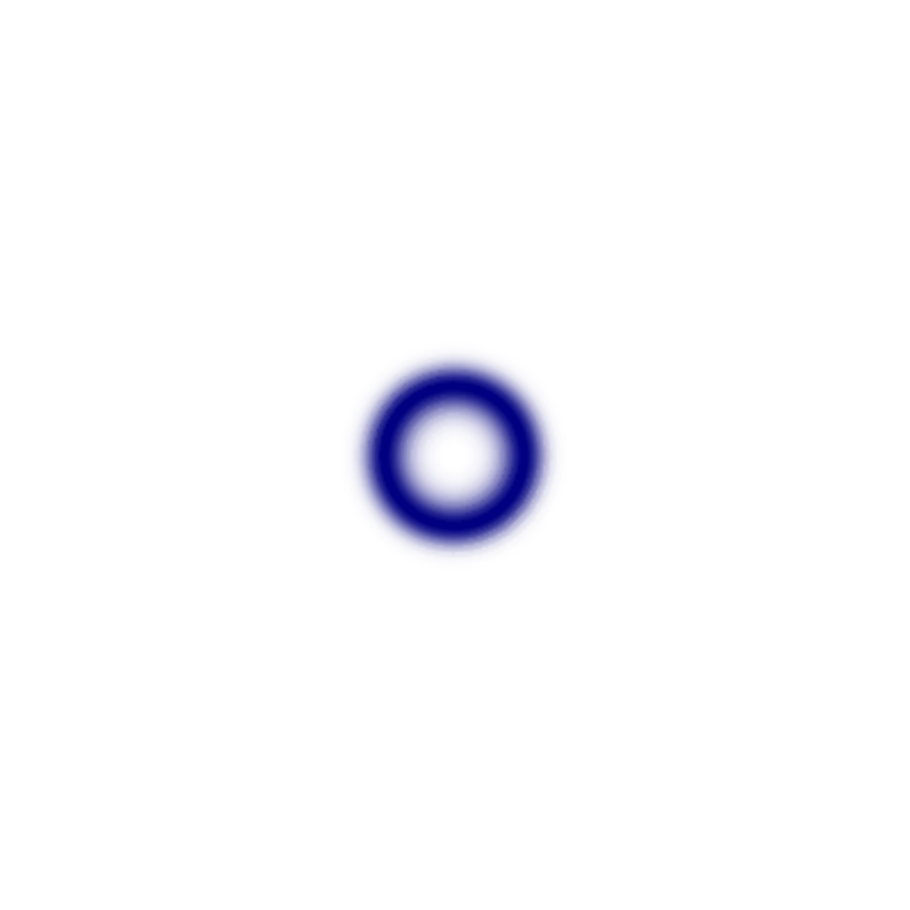}
		\caption{In the top panel, we show the magnetic field for $l=1$ (dotted line), $2,\,4,\,8,\,16,\,32$ and the limit $l\to\infty$ (dashed line). In the bottom panels, we show the magnetic field in the plane for $l=1$ (left) and for the limit $l\to\infty$ (right). The darkness of the color is related to the intensity of the magnetic field.}
		\label{figmag}
		\end{figure}
		\begin{figure}[t!]
		\centering
		\includegraphics[width=8.5cm,trim={0.65cm 1cm 0 0},clip]{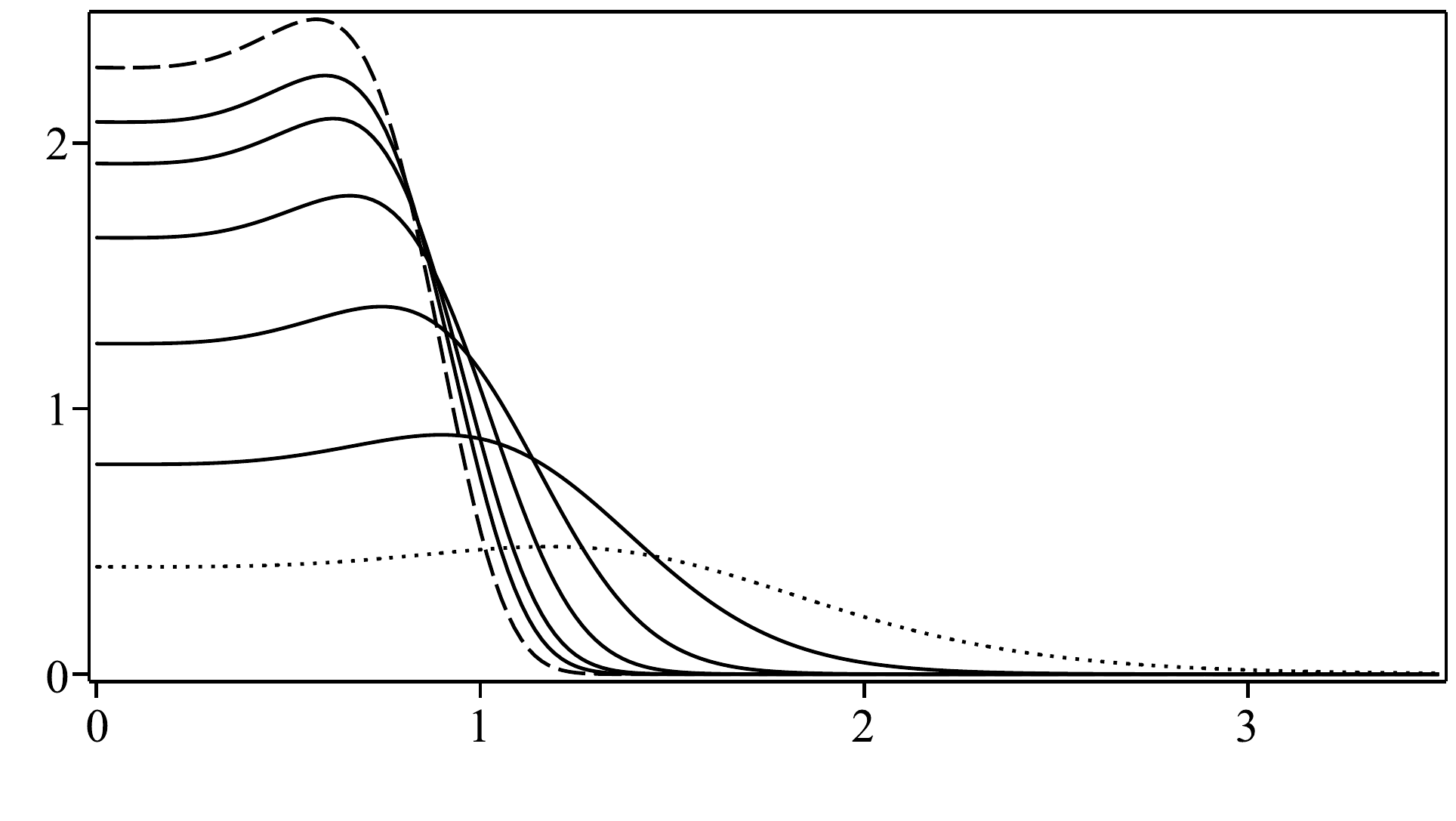}
		\includegraphics[width=4.25cm,trim={0.5cm 0.5cm 0.5cm -0.5cm},clip]{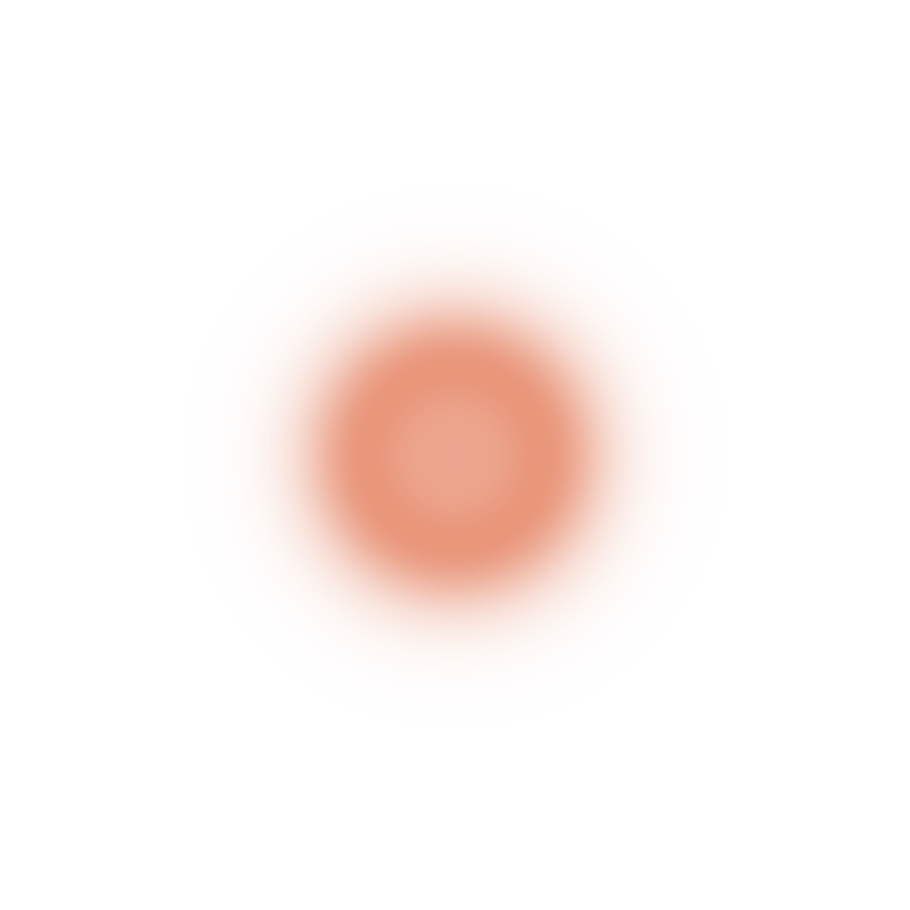}
		\includegraphics[width=4.25cm,trim={0.5cm 0.5cm 0.5cm -0.5cm},clip]{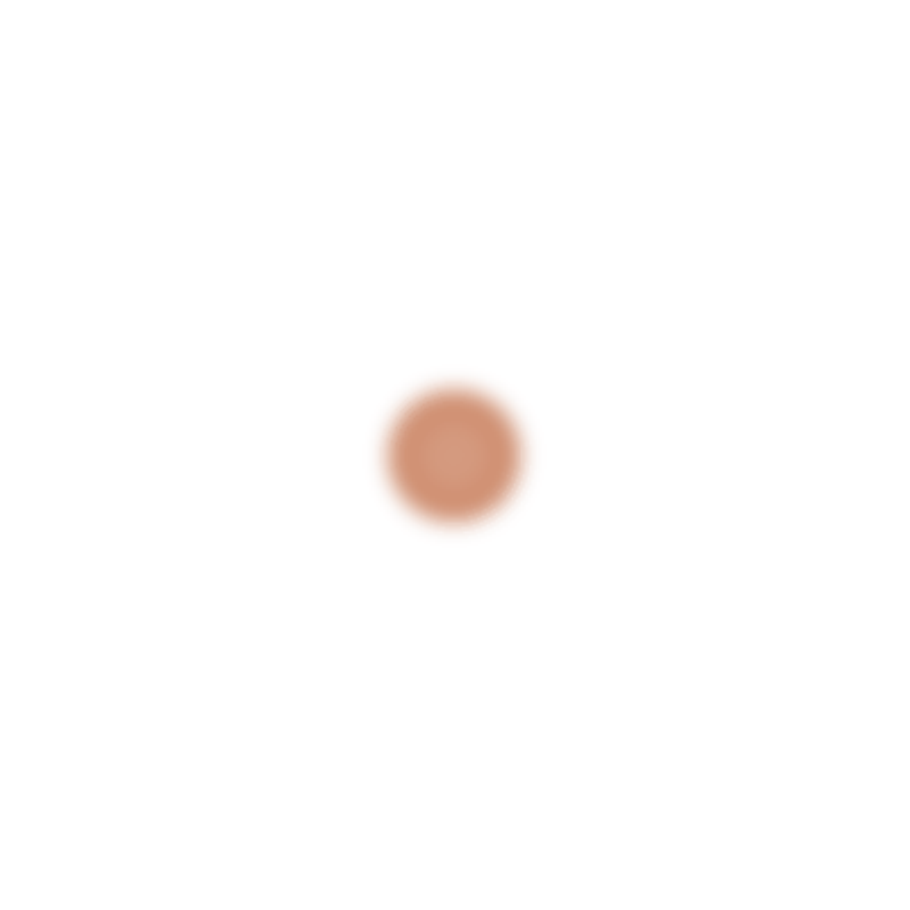}
		\caption{The energy density for $l=1$ (dotted line), $2,\,4,\,8,\,16,\,32$ and the limit $l\to\infty$ (dashed line). In the bottom panels, we show the energy density in the plane for $l=1$ (left) and for the limit $l\to\infty$ (right). The darkness of the color is related to the intensity of the energy density.}
		\label{figrho}
		\end{figure}

The magnetic field is given by Eq.~\eqref{b}; it is plotted in Fig.~\ref{figmag}. Due to the form of the first order equations Eq.~\eqref{fol} and \eqref{foinf}, the peak of the magnetic field appears when $g(r)$ attains a value that is the maximum point of the potential in $|\vphi|\in[0,1]$. The energy density comes from Eq.~\eqref{rhov}; we depict it in Fig.~\ref{figrho}. 

The planar profiles in the figures show that the magnetic field has the form of a ring and the energy density has the form of a disk. Both of them shrink as $l$ increases and we see that the blur that appear in the external boundary almost disappear when $l$ is very large; this highlights the \emph{quasi}-compact character of the vortex. In fact, one can use the asymptotic behavior in Eq.~\eqref{dexp} to show that the magnetic field and the energy density, far from the origin, have the form
\bes
\begin{align}
	B(r)&\approx \frac{\lambda^2\sigma^2}{2} e^{\sigma r}e^{-\lambda e^{\sigma r}},\\
	\rho(r)&\approx \frac{\lambda^2\sigma^4}{8} e^{2\sigma r}e^{-2\lambda e^{\sigma r}}.
\end{align}
\ees
These expressions describe the tail of $B(r)$ and $\rho(r)$, showing that they both disappear as $r$ increases due to the double exponential factor. Since the asymptotic behavior of all quantities associated to the vortex is controlled by the double exponential function, the structure describes a quasi-compact vortex.

The presence of the new quasi-compact behavior suggests that we investigate other possibilities. In particular, since the tail of these structures dies out faster than usual, one may investigate how they interact with themselves. In this case, we expect the interaction to be of a short range type. One may also exchange the Maxwell term for the Chern-Simons \cite{jackiw,coreanos} one to investigate the presence of \emph{quasi}-compact vortices with electrical charge. Another direction is to try to add an extra $U(1)$ symmetry, to accommodate additional fields that include the hidden sector in our model \cite{hidden1,hidden2}, to see how the double exponential tail modifies the hidden vortex. Since the potential in Fig.~\ref{figpot} contains the minima $|\vphi|=0,1$ and $\sqrt{2}$, it would be of interest to search for other solutions. These issues are currently under investigation and we hope to report on them in the near future.

\textit{Acknowledgments}. The work is supported by the Brazilian agencies Coordena\c{c}\~ao de Aperfei\c{c}oamento de Pessoal de N\'ivel Superior (CAPES), grant No.~88887.463746/2019-00 (MAM),  Conselho Nacional de Desenvolvimento Cient\'ifico e Tecnol\'ogico (CNPq), grants Nos. 306614/2014-6 (DB) and 306504/2018-9 (RM), and by Paraiba State Research Foundation (FAPESQ-PB) grants Nos. 0003/2019 (RM) and 0015/2019 (DB).


\end{document}